\documentclass[aps,preprint,amsmath,amssymb]{revtex4}
\usepackage{graphicx}
\def\tm{\tilde{m}}
\def\hmp{\hat m_{P}}

\begin{document}
\title{\Large \bf  $B\to \eta^{(\prime)} (\ell^{-}
\bar\nu_{\ell},\, \ell^{+} \ell^{-},\, K,\, K^*)$ decays
in the quark-flavor mixing scheme}
\date{\today}
\author{ \bf A.~G.~Akeroyd$^{1,2}$\footnote{Email: akeroyd@mail.ncku.edu.tw}
, Chuan-Hung~Chen$^{1,2}$\footnote{Email: physchen@mail.ncku.edu.tw}
and Chao-Qiang~Geng$^{3,4}$\footnote{Email: geng@phys.nthu.edu.tw} }
\affiliation{ $^{1}$Department of Physics, National Cheng-Kung
University, Tainan 701, Taiwan \\
$^{2}$National Center for Theoretical Sciences, Taiwan\\
$^{3}$Department of Physics, National Tsing-Hua University, Hsinchu
300, Taiwan  \\
$^{4}$Theory Group, TRIUMF, 4004 Wesbrook Mall, Vancouver, B.C V6T 2A3, Canada
 }

\begin{abstract}
In the quark-flavor mixing scheme, $\eta$ and $\eta'$ are linear
combinations of flavor states
$\eta_{q}=(u\bar{u}+d\bar{d})/\sqrt{2}$ and $\eta_{s}=s\bar{s}$ with
the masses of $m_{qq}$ and $m_{ss}$, respectively.
Phenomenologically, $m_{ss}$ is strictly fixed to be around $0.69$,
which is close to $\sqrt{2m^{2}_{K}-m^{2}_{\pi}}$ by the approximate
flavor symmetry, while $m_{qq}$ is found to be $0.18\pm 0.08$ GeV.
For a large allowed value of $m_{qq}$, we show that the BRs for
$B\to \eta^{(\prime)} X$ decays with $X=(\ell^{-} \bar\nu_{\ell},\;
\ell^{+} \ell^{-})$ are enhanced. We also illustrate that 
$BR(B\to\eta X)> BR(B\to \eta^{\prime} X)$ in the 
mechanism without the flavor-singlet contribution. Moreover, we demonstrate that the decay branching ratios
(BRs) for $B\to \eta^{(\prime)}K^{[*]}$ are consistent with the
data. In particular, the puzzle of the large $BR(B\to \eta^{\prime}
K)$ can be solved. In addition, we find that the CP asymmetry for
$B^{\pm}\to \eta K^{\pm}$ can be as large as $-30\%$, which agrees
well with the data. However, we cannot accommodate the CP
asymmetries of $B\to \eta K^*$ in our analysis, which could indicate
the existence of some new CP violating sources.

\end{abstract}
\maketitle

\section{Introduction}

The branching ratio (BR) of $B^0 \to \eta^{\prime} K^0$ was first
observed by the CLEO collaboration with $(89^{+18}_{-16}\pm 9)\times
10^{-6}$ \cite{CLEO_PRL81}, which is much larger than $(20-40)\times
10^{-6}$ estimated by the factorization ansatz \cite{AG_PRD}. With
more data accumulated, this incomprehensible value becomes a real
puzzle now that the measurements from BELLE and BABAR depart from
the theoretical estimations, where the former has observed
$BR(B^{+}\to \eta K^{+})=(1.9 \pm 0.3^{+0.2}_{- 0.1})\times 10^{-6}$
\cite{belle_0608}, $ BR(B^{+}\to \eta^{\prime} K^+)=(69.2\pm 2.2 \pm
3.7)\times 10^{-6}$ and $BR(B^0\to \eta^{\prime}
K^0)=(58.9^{+3.6}_{-3.5}\pm 4.3)\times 10^{-6}$ \cite{belle_0603},
while the latter has measured $BR(B^{+}\to \eta K^{+})=(3.3 \pm
0.6\pm 0.3)\times 10^{-6}$ \cite{babar_PRL95}, $BR(B^+\to
\eta^{\prime} K^+)=(68.9 \pm 2.0 \pm 3.2)\times 10^{-6}$ and
$BR(B^0\to \eta^{\prime} K^0)=(67.4\pm 3.3 \pm 3.2)\times 10^{-6}$
\cite{babar_PRL94}. To unravel the mystery, many solutions have been
proposed, such as the intrinsic charm in $\eta^{\prime}$
\cite{icharm}, the gluonium state \cite{gluon}, the spectator hard
scattering mechanism \cite{2yang} and the flavor-singlet component
in $\eta^{\prime}$ \cite{BN_NPB}. Nevertheless, there are still no
conclusive solutions yet.

Recently, the BaBar Collaboration \cite{Babar_ICHEP06} has also measured
the semileptonic decays with the data as follows:
\begin{eqnarray}
BR(B^{+}\to \eta \ell^{+} \nu_{\ell} )&=&(0.84\pm 0.27 \pm
0.21)\times
10^{-4}< 1.4\times 10^{-4} (\rm 90\%\ C.L.)\, ,\nonumber \\
BR(B^{+}\to \eta^{\prime} \ell^{+} \nu_{\ell} )&=&(0.33\pm 0.60 \pm
0.30)\times 10^{-4}< 1.3\times 10^{-4} (\rm 90\%\ C.L.)\, .
\label{Data}
\end{eqnarray}
Although the significance of the former in Eq. (\ref{Data}) is
$2.55\sigma$, the central value is a factor of 2 larger than
$0.4\times 10^{-4}$ calculated by the light-cone sum rules (LCSRs)
\cite{CG}. Due to these results, we speculate that the mechanism to
enhance the BRs of $B\to \eta^{\prime} K$ may also affect the
semileptonic decays  of $B^{-}\to \eta^{(\prime)} \ell
\bar\nu_{\ell}$. After surveying various proposed mechanisms, one
finds that only the flavor-singlet mechanism (FSM) \cite{BN_NPB}
could have direct influence on the BRs of semileptonic decays
\cite{CG,KOY}.
In this paper, inspired by the measurements of the semileptonic
decays, we would like to propose another possible mechanism within
the quark-flavor mixing scheme to study the decays of  $B\to
\eta^{(\prime)}(\ell^{-} \bar\nu_{\ell},\, \ell^{+} \ell^{-},\, K,\,
K^*)$.
We will also compare our results with those in the FSM
\cite{BN_NPB,CG,KOY,WZ,CKL} and explore the differences between the
two
 mechanisms, which could be tested in future B experiments.

The paper is organized as follows. In Sec. II, we review the
quark-flavor mixing scheme. In Sec. III, we carry out a general
analysis for the decay amplitudes and form factors. Numerical
results and discussions are presented in Sec. IV. Our conclusions
are given in Sec. V.

\section{The quark-flavor mixing scheme}

 It is known that the physical states $\eta$ and
$\eta^{\prime}$ are composed of the flavor octet $\eta_{8}$ and
singlet $\eta_{1}$, in which the flavor wave functions are denoted
as $\eta_{8}=(u\bar{u}+d\bar{d}-2s\bar{s})/\sqrt{6}$ and
$\eta_{1}=(u\bar{u}+d\bar{d}+s\bar{s})/\sqrt{3}$, respectively. Due
to the $U_{A}(1)$ anomaly, it is understood that the mass of
$\eta^{\prime}$ is much larger than that of $\eta$. To satisfy the
current experimental data, usually one needs to introduce two angles
to the mixing matrix, defined by $\eta=\cos \theta _8 \eta_8-\sin
\theta _1 \eta_1$ and $\eta^{\prime}=\sin \theta _8 \eta_8+\cos
\theta _1 \eta_1$ \cite{Leutwyler,AG_PRD},
to describe the connection between physical and flavor states.
However, it is known that by using the two-angle scheme, we will
encounter a divergent problem in some $B$ decays \cite{Ali}, such as
$B\to\eta'K$. To illustrate this problem, we notice that in these
decays, the factorized parts
are associated with the matrix element $\langle 0| \bar{s}
i\gamma_{5} s| \eta_{1}\rangle$.  From the equation of motion, one
has $\langle 0 |\partial ^\mu \bar s\gamma _\mu  \gamma _5 s|
{\eta_{1}} \rangle  = \langle 0 |2m_s \bar si\gamma _5 s| {\eta_{1}}
\rangle= m_{\eta_{1}}^2 f_{\eta_{1}} $, leading to $\langle 0 |\bar
s i\gamma _5 s| \eta_{1} \rangle=m_{\eta_{1}}^2f_{\eta_{1}}/2m_s$,
 where $f_{\eta_{1}}(m_{\eta_{1}})$
is the decay constant (mass) of $\eta_{1}$.
In the chiral limit of $m_{s}\to 0$, the matrix element diverges because
$m_{\eta_{1}}\neq 0$.
To explicitly display the chiral limit, it is better to use
the quark-flavor scheme,
defined by \cite{flavor0,flavor}
\begin{eqnarray}
\left( {\begin{array}{*{20}c}
   \eta   \\
   {\eta '}  \\
\end{array}} \right) = \left( {\begin{array}{*{20}c}
   {\cos \phi } & { - \sin \phi }  \\
   {\sin \phi } & {\cos \phi }  \\
\end{array}} \right)\left( {\begin{array}{*{20}c}
   {\eta _{q} }  \\
   {\eta _{s} }  \\
\end{array}} \right) \,,\label{eq:flavor}
\end{eqnarray}
where $\eta _{q}  = ( {u\bar u + d\bar d})/\sqrt{2}$ and $\eta_{s} =
s\bar s $.
 From the definition of $\langle 0| \bar q'
\gamma_{\mu} \gamma_{5} q'|
\eta_{q'}(p)\rangle=if_{\eta_{q'}}p_{\mu}$ ($q'=q,s$), the masses of
$\eta_{q,s}$ can be expressed by
\begin{eqnarray}
m_{qq}^2  &=& \frac{\sqrt 2}{f_{\eta_q} }\langle 0|m_u \bar
ui\gamma _5 u + m_d \bar di\gamma _5 d| \eta_q \rangle,\ \ \
m_{ss}^2  = \frac{2}{f_{\eta_s}}\langle 0 |m_s \bar si\gamma _5 s|
\eta _s \rangle. \label{eq:masses}
\end{eqnarray}
Clearly, in terms of the quark-flavor basis, $m_{qq}$ and $m_{ss}$ are
 zero in the chiral limit. We note that
$m_{qq}$ and $m_{ss}$ are unknown parameters and their values can
be obtained by fitting with the data, such as the masses of
$\eta^{(\prime)}$ and the decay rates of some relevant $B$ decays.
Note that $m_{qq,ss}$
  are related to $m^{0}_{\eta_q,\eta_s,K}$
 by
$m^{0}_{\eta_q}=m^{2}_{qq}/(m_u+m_d)$,
$m^{0}_{\eta_s}=m^2_{ss}/2m_{s}$ and
$m^{0}_{K}=m^{2}_{K}/(m_s+m_q)$.
 From the divergences of the
 axial vector currents
\begin{eqnarray}
\partial ^\mu  \bar q^{\prime} \gamma_{\mu}\gamma_5 q^{\prime}
  &=& \frac{ \alpha _s }{4\pi }G
\tilde G +  2 m_{q^{\prime}} \bar{q}^{\prime}i\gamma _5 q^{\prime}
,
\label{eq:axial}
\end{eqnarray}
where $G=G^{a\mu\nu}$ are the gluonic field-strength and $\tilde
G=\tilde G^{a\mu\nu}\equiv\epsilon^{\mu\nu\alpha
\beta}G^{a}_{\alpha\beta}$,
one obtains the $\eta_{q,s}$ masses as
\begin{eqnarray}
\left( {\begin{array}{*{20}c}
   {M_{qq}^2 } & {M_{qs}^2 }  \\
   {M_{sq}^2 } & {M_{ss}^2 }  \\
\end{array}} \right) & = &\left( {\begin{array}{*{20}c}
   \langle 0|\partial^\mu  J_{\mu 5}^q | {\eta _q }\rangle /f_q  & \langle 0 |\partial^\mu  J_{\mu 5}^s | \eta _q \rangle /f_s  \\
   \langle 0|\partial^\mu  J_{\mu 5}^q | {\eta _s } \rangle /f_q & \langle 0 |\partial^\mu  J_{\mu 5}^s | \eta _s  \rangle/f_s   \\
\end{array}} \right) \nonumber \\
  & = & \left( {\begin{array}{*{20}c}
   m_{qq}^2  + 2a^2 & \sqrt{2} y a^2  \\
   \sqrt{2} y a^2 & m_{ss}^2  + y^2 a^2       \\
\end{array}} \right)
\end{eqnarray}
with $ a^2 = \langle 0| \alpha_s G \tilde
G|\eta_q\rangle/(4\sqrt{2}\pi f_{q})$ and $y=f_{q}/f_{s}$.
 Furthermore, by using the mixing matrix introduced
in Eq.~(\ref{eq:flavor}),
we have \cite{FK}
\begin{eqnarray}
\sin\phi &=&
\left[\frac{(m^{2}_{\eta^{\prime}}-m^2_{ss})(m^2_{\eta}-m^2_{qq})}{(m^{2}_{\eta^{\prime}}-m^2_{\eta})(m^2_{ss}-m^2_{qq})}
\right]^{1/2}, \nonumber \\
y &=&
\left[2\frac{(m^{2}_{\eta^{\prime}}-m^2_{ss})(m^2_{ss}-m^2_{\eta})}{(m^{2}_{\eta^{\prime}}-m^2_{qq})(m^2_{\eta}-m^2_{qq})}
\right]^{1/2}, \nonumber \\
a^2&=&\frac{1}{2}\frac{(m^{2}_{\eta^{\prime}}-m^2_{qq})(m^2_{\eta}-m^2_{qq})}{m^2_{ss}-m^2_{qq}},
\label{eq:parameters}
\end{eqnarray}
where $m_{\eta^{(\prime)}}$ is the mass of $\eta^{(\prime)}$.

According to the relations in Eq.~(\ref{eq:masses}),
it is interesting to see
 that the parameter $m_{qq(ss)}$ is involved in the
distribution amplitude of the $\eta_{q(s)}$ state, which is defined by
\cite{KLS_PRD65}
\begin{eqnarray}
\langle 0 | \bar q''(0)_{j} q''_{k}(z)|
\eta_{q^{\prime}}(p)\rangle & = &\frac{i}{\sqrt{2N_c}}\int^{1}_{0}
dx e^{-ixp\cdot z} [ (\not p \gamma_{5})_{jk} \phi_{\eta_{q'}}(x)
\nonumber
\\
&&+ (\gamma_{5})_{jk} m^{0}_{\eta_{q'}} \phi^{\rho}_{\eta_{q'}}(x)
+ m^{0}_{\eta_{q'}}[\not n_{-} \not n_{+} -1]_{jk}
\phi^{\sigma}_{\eta_{q'}}(x)]\,, \label{eq:da}
\end{eqnarray}
where $q''=u, d$ and $s$, $q^{\prime}=q$ and $s$,
$\phi_{\eta_{q'}}(x)$ and $\phi^{\rho(\sigma)}_{\eta_{q'}}(x)$
denote the twist-2 and twist-3 wave functions of the $\eta_{q'}$
state, respectively, $x$ is the momentum fraction,
$m^{0}_{\eta_{q'}}$ stands for the chiral symmetry breaking
parameter, and $n_{+}=(1,0,\textbf{0})$ and $n_{-}=(0,1,
\textbf{0})$ are defined in the light-cone coordinates. On
substituting Eq.~(\ref{eq:da}) into Eq.~(\ref{eq:masses}), we obtain
$m^{0}_{\eta_{q}}=m^{2}_{qq}/(m_{u}+m_{d})$ and
$m^{0}_{\eta_s}=m^2_{ss}/2m_s$. In the next section, it will be
clear that
 the value of $m_{qq}$ is crucial
for the determination of the $B\to \eta^{(\prime)}$ transition form
factors, which play important roles in
the decay branching ratios of $B\to
\eta^{(\prime)} X$ with $X=(\ell^{-} \bar\nu_{\ell},\,\ell^{+}
\ell^{-},\, K)$.

\section{Decay amplitudes and form factors}

We first study the semileptonic decays of $B^{-}\to \eta^{(\prime)}\ell^{-} \nu_{\ell}$ and
$\bar B\to \eta^{(\prime)}\ell^{+} \ell^{-}$ by writing the effective
Hamiltonians  at quark level in the
SM  as
\begin{eqnarray}
{\cal H}_{I}&=& \frac{G_FV_{ub}}{\sqrt{2}}  \bar u\gamma_{\mu}
(1-\gamma_5) b\, \bar \ell \gamma^{\mu} (1-\gamma_5) \nu_{\ell}\,,
\label{eq:heff_lnu}
\\
 {\cal H}_{II} &=& \frac{G_F\alpha_{em} \lambda^{q^{\prime}}_t}{\sqrt{2}
 \pi}\left[ H_{1\mu} L^{\mu} +H_{2\mu}L^{5\mu}  \right]\,,
 \label{eq:heff_ll}
  \end{eqnarray}
respectively,
  with
  \begin{eqnarray}
  H_{1\mu } &=&C^{\rm eff}_{9}(\mu )\bar{q}^{\prime}\gamma _{\mu }P_{L}b\ -\frac{2m_{b}}{%
 q^{2}}C_{7}(\mu )\bar{q}^{\prime}i\sigma _{\mu \nu }q^{\nu }P_{R}b \,,
\nonumber \\
 H_{2\mu } &=&C_{10}\bar{q}^{\prime}\gamma _{\mu }P_{L}b \,,
 \nonumber\\
 L^{\mu } &=&\bar{\ell}\gamma ^{\mu }\ell\,, \ \ \ L^{5\mu } =\bar{\ell}\gamma ^{\mu }\gamma
 _{5}\ell\,,
  \label{heffc}
  \end{eqnarray}
where $\alpha_{em}$ is the fine structure constant, $V_{ij}$ denote
the Cabibbo-Kobayashi-Maskawa (CKM) matrix elements,
$\lambda^{q^{\prime}}_t=V_{tb}V_{tq^{\prime}}^*$, $C_{i}$ are the
Wilson coefficients (WCs) with their explicit expressions given in
Ref.~\cite{BBL},
$m_b$ is the current b-quark mass, $q$ is the momentum transfer
and $P_{L(R)}=(1\mp \gamma_5)/2$. Note that the long-distance
effects of $c\bar{c}$ bound states have been included in $C^{\rm
eff}_9$, given by \cite{CG_PRD66}
\begin{eqnarray}
C_{9}^{\rm eff}(\mu)&=&C_{9}( \mu ) +\left( 3C_{1}\left( \mu
\right) +C_{2}\left( \mu \right) \right) \left( h\left( z,s\right)
-\frac{3}{\alpha^{2}_{em}}\sum_{V=\Psi ,\Psi ^{\prime
}}k_{V}\frac{\pi \Gamma \left( V\rightarrow
\ell^{+}\ell^{-}\right)
M_{V}}{M_{V}^{2}-q^{2}-iM_{V}\Gamma _{V}}\right) \,, \nonumber \\
h(z,s)&=&-\frac{8}{9}\ln\frac{m_b}{\mu}-\frac{8}{9}\ln z
+\frac{8}{27} +\frac{4}{9}x  -\frac{2}{9}(2+x)|1-x|^{1/2}
\nonumber
\\
&\times& \left\{
  \begin{array}{c}
    \ln \left|\frac{\sqrt{1-x}+1}{\sqrt{1-x}-1} \right|-i\, \pi, \  {\rm for\ x\equiv 4z^2/s<1 }\, , \\
    2\, arctan\frac{1}{\sqrt{x-1}},\   {\rm for\ x\equiv 4z^2/s>1 }  \, ,\\
  \end{array}
\right.
\end{eqnarray}
where $h(z,s)$ describes the one-loop matrix elements of operators
$O_{1}= \bar{s}_{\alpha }\gamma ^{\mu }P_{L}b_{\beta }\
\bar{c}_{\beta }\gamma _{\mu }P_{L}c_{\alpha }$ and
$O_{2}=\bar{s}\gamma ^{\mu }P_{L}b\ \bar{c}\gamma _{\mu }P_{L}c$
\cite{BBL} with $z=m_c/m_b$ and $s=q^2/m^2_b$, $M_{V}$ ($\Gamma
_{V}$) are the masses (widths) of intermediate states. The
hadronic matrix elements for
the  $B\to P$ transition are parametrized  as
\begin{eqnarray}
\langle P(p_{P}) | \bar q^{\prime} \gamma^{\mu}  b| \bar
B(p_B)\rangle &=& f^{P}_{+}(q^2)\left(P^{\mu}-\frac{P\cdot
q}{q^2}q^{\mu} \right)+f^{P}_{0}(q^2) \frac{P\cdot q}{q^2} q_{\mu}\,
, \nonumber
\\
\langle P(p_{P} )| \bar q^{\prime} i\sigma_{\mu\nu} q^{\nu}b| \bar B
(p_{B})\rangle &=& {f^{P}_{T}(q^2)\over m_{B}+m_{P}}\left[P\cdot q\,
q_{\mu}-q^{2}P_{\mu}\right] \label{eq:bpff}
\end{eqnarray}
with $P$ representing the pseudoscalar, $P_{\mu}=(p_B+p_P)_{\mu}$,
$q_{\mu}=(p_B-p_P)_{\mu}$ and $f_{\alpha}^{P}(q^{2})$ are form
factors.  Consequently, the transition amplitudes associated with
the interactions in Eqs.~(\ref{eq:heff_lnu}) and (\ref{eq:heff_ll})
can be expressed as
 \begin{eqnarray}
       {\cal M}_{I}&=&\frac{\sqrt{2}G_{F}V_{ub}}{\pi }
        f^{P}_{+}(q^2) \bar{\ell} \not{p}_{P} \ell\, ,
       \label{amppln}
       \\
       {\cal M}_{II}&=&\frac{G_{F}\alpha_{em} \lambda^{q^{\prime}} _{t}}{\sqrt{2}\pi }
       \left[ \tm_{97} \bar{\ell} \not{p}_P \ell + \tm_{10} \bar{\ell} \not{p}_P \gamma_5 \ell
        \right]\label{amppll}
 \end{eqnarray}
for $\bar B\rightarrow P \ell^{-} \bar\nu_{\ell}$ and
  $\bar B\rightarrow P \ell^{+} \ell^{-}$, respectively,
 where
 \begin{eqnarray}
  \tm_{97}&=& C^{\rm eff}_9 f^{P}_+(q^2) +\frac{2m_b}{m_B+m_{P}}C_7
  f^{P}_T(q^2) \,, \ \ \  \tm_{10}= C_{10} f^{P}_+(q^2)\, .
  \label{eq:m7910}
 \end{eqnarray}
The differential decay rates for $ B^{-}\to P \ell^{-}
\bar\nu_{\ell}$ and $\bar B_{d} \to P \ell^+ \ell^-$ as  functions
of $q^2$ are given by \cite{CG}
\begin{eqnarray}
\frac{d\Gamma_{I}}{dq^2 }&=& \frac{G^{2}_{F} |V_{ub}|^2
m^3_{B}}{3\cdot 2^6 \pi^3}\sqrt{(1-s+\hmp^2)^2-4\hmp^2}
\left(f^{P}_{+}(q^2) \hat P_{P}\right)^2\,,
 \label{eq:diffplnu}
 \\
\frac{d\Gamma _{II} }{dq^2
}&=&\frac{G_{F}^{2}\alpha^{2}_{em}m^{3}_{B}}{ 3\cdot 2^{9} \pi
^{5}} |\lambda^{q^{\prime}} _{t}|^{2}\sqrt{(1-s+\hmp^2)^2-4\hmp^2}
\hat P^2_{P} \left( |\tm_{97}|^2+|\tm_{10}|^2\right)
\label{eq:difpll}\, ,
 \end{eqnarray}
respectively, with $\hat P_{P}=2\sqrt{s}
|\vec{p}_{P}|/m_{B}=\sqrt{(1-s-\hmp^2 )^2-4s\hmp^2}$. Since we
concentrate on the production of the light leptons, we have
neglected the terms explicitly related to the lepton mass. We note
that due to $C_9>>C_7$, the effect associated with the form factor
of $f^{P}_{T}(q^2)$ in Eq.~(\ref{eq:m7910}) is small. From
Eqs.~(\ref{eq:diffplnu}) and (\ref{eq:difpll}), we see clearly that
the semileptonic decays are only sensitive to the form factor
$f^{P}_{+}(q^2)$. By this property, we can use the data of $B^{-}\to
\eta \ell^{-}\bar\nu_{\ell}$ to constrain the unknown parameters in
the calculations of the form factors. The constrained parameters
could make some predictions for the decays $B\to
\eta^{(\prime)}\ell^{+}\ell^{-}$ and $B\to \eta^{(\prime)} K$.

In the large recoil region, i.e. $q^{2}\to 0$, the form
factors can be evaluated  by the perturbative QCD (PQCD)
\cite{PQCD1,PQCD2} approach, in which the transverse momenta of
valence quarks are included to remove the end-point singularities
when $x\to 0$.
\begin{figure}[htbp]
\includegraphics*[width=4.5in]{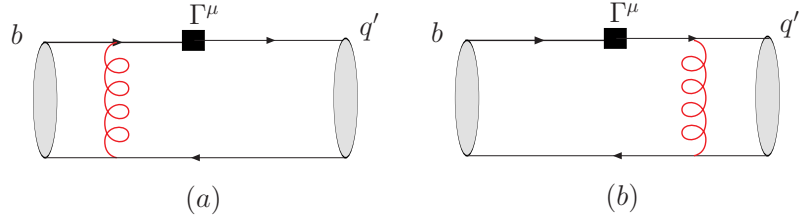}
\caption{Flavor diagrams for the $B\to P$ transition
with $\Gamma^{\mu}=(\gamma^{\mu},\, i\sigma^{\mu\nu}q_{\nu})$.}
 \label{fig:ff}
\end{figure}
Hence, in terms of Eq.~(\ref{eq:da}) and the flavor diagrams
shown in Fig.~\ref{fig:ff},
the form factors $f^{P}_{+}(q^2)$, $f^{P}_{-}(q^2)$,
and $f^{P}_{T}(q^2)$ for $B\rightarrow P$ can be formulated as
\cite{CG_PRD66}
\begin{eqnarray}
f^{P}_{+}(q^2)&=&f^{P}_1(q^2) + f^{P}_2 (q^2)\,,
 \nonumber \\
f^{P}_{0}(q^2)&=&f^{P}_1(q^2)
\Big(1+\frac{q^2}{m^2_{B}}\Big)+f^{P}_2(q^2)
\Big(1-\frac{q^2}{m^2_{B}}\Big)\,, \label{eq:f0}
\end{eqnarray}
where
\begin{eqnarray}
f^{P}_{1}( q^{2} ) &=&8\pi
C_{F}m_{B}^{2}r_{P}\int_{0}^{1}[dx]\int_{0}^{\infty
}b_{1}db_{1}b_{2}db_{2}\ \phi _{B}( x_{1},b_{1})  \left[
\phi^{p}_{P}(x_2)-\phi^{t}_{P}(x_{2}) \right]\nonumber \\
&&\times E(t^{(1)})h( x_{1},x_{2},b_{1},b_{2})\,,
 \nonumber \\
f^{P}_{2}( q^{2} ) &=&8\pi
C_{F}m_{B}^{2}\int_{0}^{1}[dx]\int_{0}^{\infty
}b_{1}db_{1}b_{2}db_{2}\ \phi
_{B}( x_{1},b_{1}) \   \nonumber \\
&&\times \left\{ \left[ (1+x_{2}\xi)\phi_{P}(x_{2})+2r_{P}\left(
\left(\frac{1}{\xi}-x_{2}\right)\phi^{t}_{P}(x_{2})-x_{2}\phi^{p}_{P}(x_{2})\right)
\right]\right.\nonumber \\
&& \left.\times E(t^{(1)}) h(
x_{1},x_{2},b_{1},b_{2})+2r_{P}\phi^{p}_{P}(x_2)E(t^{(2)})h(
x_{2},x_{1},b_{2},b_{1})\right\}\,,
\end{eqnarray}
and
\begin{eqnarray}
f^{P}_{T}( q^{2}) &=&8\pi
C_{F}m^{2}_{B}(1+m_{P}/m_{B})\int_{0}^{1}[dx]\int_{0}^{\infty
}b_{1}db_{1}b_{2}db_{2}\phi _{B}( x_{1},b_{1})  \nonumber \\
&&\times \left\{ \left[ \phi _{P}( x_{2})-r_{P}x_{2}\phi _{P}^{p
}(x_2)+r_{P}\left(\frac{2}{\xi}+x_2\right)\phi _{P}^{t
}(x_2)\right] E( t^{( 1) })
h( x_{1},x_{2},b_{1},b_{2}) \right.\nonumber \\
&&\left. + 2r_{P}\phi _{P}^{p}( x_{2})  E( t^{( 2) }) h(
x_{2},x_{1},b_{2},b_{1}) \right\}\,,  \label{eq:ft}
\end{eqnarray}
with $C_{F}=4/3$, $\xi=1-q^2/m^{2}_{B}$ and $r_{P}=m^{0}_{P}/m_{B}$.
>From Eq. (\ref{eq:f0}), we find that $f_{+}(0)=f_{0}(0)$. The
evolution factor is given by $E( t) =\alpha _{s}( t) \exp ( -S_{B}(
t) -S_{P}( t) )$ where the Sudakov exponents $S_{B(P)}$ can be found
in Ref. \cite{ChenLi_PRD63}. The hard function $h$ is written as
\begin{eqnarray}
h(x_{1},x_{2},b_{1},b_{2}) &=&S_{t}(x_{2})K_{0}( \sqrt{x_{1}x_{2}\xi
}
m_{B}b_{1})  \nonumber \\
&&\times [ \theta (b_{1}-b_{2})K_{0}( \sqrt{x_{2}\xi } m_{B}b_{1})
I_{0}( \sqrt{x_{2}\xi }m_{B}b_{2})  \nonumber
\\ && +\theta (b_{2}-b_{1})K_{0}( \sqrt{x_{2}\xi}
m_{B}b_{2}) I_{0}( \sqrt{x_{2}\xi} m_{B}b_{1}) ]\,, \label{dh}
\end{eqnarray}
where the threshold resummation effect is described by
$S_{t}(x)=2^{1+2c}\Gamma(\frac{3}{2}+c)[x(1-x)]^{c}/
\sqrt{\pi}\Gamma(1+c)$ with $c=0.3$ \cite{PQCD2}. The hard scales $t^{(1,2)}$ are
chosen to be \cite{Chen}
\begin{eqnarray*}
t^{( 1) } &=&\max ( \sqrt{m_{B}^{2}\xi x_{2}}
,1/b_{1},1/b_{2},\bar\Lambda) \,,  \nonumber \\
t^{( 2) } &=&\max ( \sqrt{m_{B}^{2}\xi
x_{1}},1/b_{1},1/b_{2},\bar\Lambda)\,, \label{tscale}
\end{eqnarray*}
where $\bar\Lambda$ is used to exclude the effects from
nonperturbative contributions. To get the BRs for the three-body
semileptonic decays, besides the values of the form factors at
$q^{2}=0$, we also need to know  their $q^{2}$ dependences.
To obtain them,
 we adopt the
fitting results calculated by the light-cone sum rules (LCSR)
\cite{LCSR}, given by
\begin{eqnarray}
f^{P}_{+(T)}(q^2)&=&
\frac{f^{P}_{+(T)}(0)}{(1-q^2/m^2_{B^*})(1-\alpha_{+(T)}
q^2/m^2_{B^*})} \label{eq:q2_ff}
%
\end{eqnarray}
with $\alpha_{+(T)}=0.52(0.84)$ and $m_{B^*}=5.32$ GeV. In terms of
the quark-flavor mixing scheme, we will calculate the $B\to \eta_{q,s}$ form
factors, which are related to those of
$B\to\eta^{(\prime)}$  by
\begin{eqnarray}
f^{\eta}_{+(T)}(q^2)&=& \frac{\cos\phi}{\sqrt{2}}
f^{\eta_q}_{+(T)}(q^2)\,, \nonumber \\
f^{\eta^{\prime}}_{+(T)}(q^2)&=& \frac{\sin\phi}{\sqrt{2}}
f^{\eta_q}_{+(T)}(q^2)\,. \label{eq:eta_ff}
\end{eqnarray}

\begin{figure}[htbp]
\includegraphics*[width=4.5in]{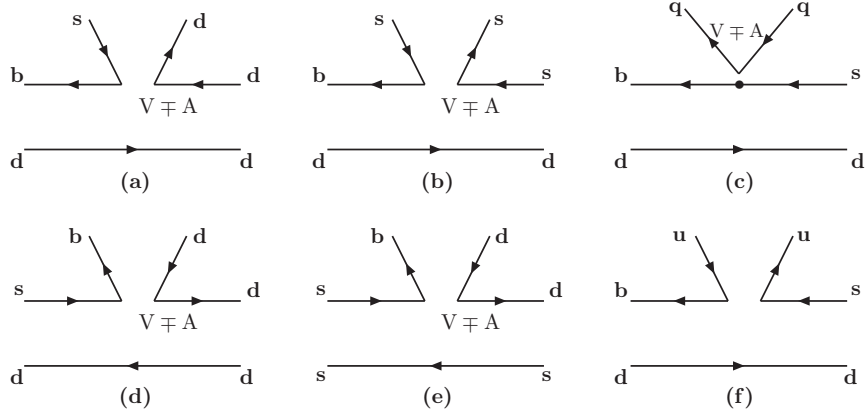}  \caption{Flavor diagrams for $B_{d}\to \eta^{(\prime)} K^{0}$
decays: (a)-(e) stand for the penguin contributions while (f) is the
tree contribution, where $V\mp A$ denote the left-hand and right-handed
currents, respectively.}
 \label{fig:flavor}
\end{figure}

\begin{figure}[htbp]
\includegraphics*[width=4.5in]{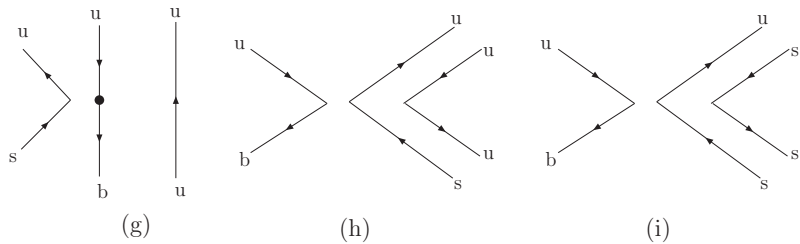}  \caption{Flavor diagrams arising from
tree emission and annihilation for charged $B$ decays.}
 \label{fig:tree}
\end{figure}

For the nonleptonic decays of $B\to \eta^{(\prime)} K$,
we will assume the color-transparency
\cite{Bjorken}, $i.e.$, no rescattering effects in $B$
decays. The effective interaction
for the $b\to s$ transition at the quark level is given by
\cite{BBL}
\begin{equation}
H_{{\rm eff}}={\frac{G_{F}}{\sqrt{2}}}\sum_{q=u,c}V_{q}\left[
C_{1}(\mu) O_{1}^{(q)}(\mu )+C_{2}(\mu )O_{2}^{(q)}(\mu
)+\sum_{i=3}^{10}C_{i}(\mu) O_{i}(\mu )\right] \;,
\label{eq:hamiltonian}
\end{equation}
where $V_{q}=V_{qs}^{*}V_{qb}$ are the CKM matrix elements and the
operators $O_{1}$-$O_{10}$ are defined as
\begin{eqnarray}
&&O_{1}^{(q)}=(\bar{q}^{\prime}_{\alpha}q_{\beta})_{V-A}(\bar{q}_{\beta}b_{\alpha})_{V-A}\;,\;\;\;\;\;
\;\;\;O_{2}^{(q)}=(\bar{q}^{\prime}_{\alpha}q_{\alpha})_{V-A}(\bar{q}_{\beta}b_{\beta})_{V-A}\;,
\nonumber \\
&&O_{3}=(\bar{q}^{\prime}_{\alpha}b_{\alpha})_{V-A}\sum_{q}(\bar{q}_{\beta}q_{\beta})_{V-A}\;,\;\;\;
\;O_{4}=(\bar{q}^{\prime}_{\alpha}b_{\beta})_{V-A}\sum_{q}(\bar{q}_{\beta}q_{\alpha})_{V-A}\;,
\nonumber \\
&&O_{5}=(\bar{q}^{\prime}_{\alpha}b_{\alpha})_{V-A}\sum_{q}(\bar{q}_{\beta}q_{\beta})_{V+A}\;,\;\;\;
\;O_{6}=(\bar{q}^{\prime}_{\alpha}b_{\beta})_{V-A}\sum_{q}(\bar{q}_{\beta}q_{\alpha})_{V+A}\;,
\nonumber \\
&&O_{7}=\frac{3}{2}(\bar{q}^{\prime}_{\alpha}b_{\alpha})_{V-A}\sum_{q}e_{q} (\bar{q}%
_{\beta}q_{\beta})_{V+A}\;,\;\;O_{8}=\frac{3}{2}(\bar{q}^{\prime}_{\alpha}b_{\beta})_{V-A}
\sum_{q}e_{q}(\bar{q}_{\beta}q_{\alpha})_{V+A}\;,  \nonumber \\
&&O_{9}=\frac{3}{2}(\bar{q}^{\prime}_{\alpha}b_{\alpha})_{V-A}\sum_{q}e_{q} (\bar{q}%
_{\beta}q_{\beta})_{V-A}\;,\;\;O_{10}=\frac{3}{2}(\bar{q}^{\prime}_{\alpha}b_{\beta})_{V-A}
\sum_{q}e_{q}(\bar{q}_{\beta}q_{\alpha})_{V-A}\;, \label{eq:ops}
\end{eqnarray}
with $\alpha$ and $\beta$ being the color indices. In Eq.
(\ref{eq:hamiltonian}), $O_{1}$-$O_{2}$ are from the tree level of
weak interactions, $O_{3}$-$O_{6}$ are the so-called gluon penguin
operators and $O_{7}$-$O_{10}$ are the electroweak penguin
operators, while $C_{i}$ ($i=1,2,\cdots,10$) are the corresponding WCs. Using
the unitarity condition, the CKM matrix elements for the penguin
operators $O_{3}$-$O_{10}$ can also be expressed as
$V_{u}+V_{c}=-V_{t}$. To study the nonleptonic decays, we will
encounter the transition matrix elements such as $\langle P_{1}
P_{2}| H_{\rm eff} |B\rangle= \langle P_{1} P_{2}| V_{q} C_{i}
O_{i} |B\rangle$. To describe the B decay amplitudes,
we have to know  not only the relevant effective weak interactions
but also all possible topologies for the specific process.
In Fig. \ref{fig:flavor},
we display the flavor diagrams for $B_{d}\to \eta_{q(s)}
K$ decays,
in which (a)-(c), (d)-(e) and (f) illustrate  penguin emission,
penguin annihilation and
tree emission topologies, respectively. Since the b-quark is
dictated by the weak charged current, its chirality is always
left-handed. However, the chiralities for $q\bar{q}$ pairs, produced
by gluon, Z-boson and photon penguins, could be both left and
right-handed, resulting in processes containing both $V-A$ and $V+A$
currents.
In Fig. \ref{fig:flavor}, we have explicitly labeled
the associated type of currents
except the diagram (f)
which is from the tree and only has the left-handed interaction.
Note that although we use the
states $\eta_{q,s}$ as our basis, the physical states can be easily
obtained by using Eq.~(\ref{eq:flavor}). For the charged $B$ decays,
besides the flavor diagrams displayed in Fig.~\ref{fig:flavor},
three more diagrams arising from tree emission and annihilation
topologies need to be included as shown in Fig.~\ref{fig:tree}.
 From Figs.~\ref{fig:flavor} and \ref{fig:tree}, the decay amplitudes for $B^{0,+}\to
\eta_{q} K^{(*)0,+}$ and $B\to \eta_{s} K^{(*)0,+}$ are given by
\begin{eqnarray}
A^{0}_{q}&=&V_{t} \left( F^{0}_{Pa} + N^{0}_{Pa}+ F^{0}_{Pc} +
N^{0}_{Pc}+ F^{0}_{Pd} + N^{0}_{Pd}\right) -V_{u}
\left(F^{0}_{Tf} + N^{0}_{Tf} \right), \nonumber \\
A^{0}_{s}&=&V_{t}\left( F^{0}_{P(b+c)} + N^{0}_{P(b+c)}+F^{0}_{Pe} +
N^{0}_{Pe}\right)\label{eq:nonlep_n}
\end{eqnarray}
and
\begin{eqnarray}
A^{+}_{q}&=&V_{t} \left( F^{+}_{Pa} + N^{+}_{Pa}+ F^{+}_{Pc} +
N^{+}_{Pc}+ F^{+}_{Pd} + N^{+}_{Pd}\right) \nonumber \\
&& -V_{u}\left(F^{+}_{T(f+g)} + N^{+}_{T(f+g)}+ F^{+}_{Th} + N^{+}_{Th} \right), \nonumber \\
A^{+}_{s}&=&V_{t}\left( F^{+}_{P(b+c)} + N^{+}_{P(b+c)}+F^{+}_{Pe} +
N^{+}_{Pe}\right)-V_{u} \left( F^{+}_{Ti} + N^{+}_{Ti}\right)\,,
\label{eq:nonlep_c}
\end{eqnarray}
where $V_{t}=V_{tb}V^{*}_{ts}=-A\lambda^2$ and
$V_{u}=V_{ub}V^*_{us}=A\lambda^{4}R_{b}e^{-i\phi_{3}}$,
$F^{0,+}_{Pk}$ and $N^{0,+}_{Pk}$ represent the penguin factorized
and nonfactorized contributions for the topology $k$, and
$F^{0,+}_{Tk}$ and $N^{0,+}_{Tk}$ are the tree factorized and
nonfactorized effects, respectively. The lengthy formulas for
various factorizable and nonfactorizable parts can be found in
Refs.~\cite{ChenLi_PRD63,ES}. We note that for simplicity we have
used the same notations for the $\eta_{q,s} K$ and $\eta_{q,s} K^*$
modes. Furthermore, from Eq.~(\ref{eq:flavor}) the physical decays
can be written as
\begin{eqnarray}
A(B^{0,+}\to \eta K^{(*)})&=&\frac{\cos\phi}{\sqrt{2}} A^{0,+}_{q}
-\frac{\sin\phi}{\sqrt{2}} A^{0,+}_{s}, \nonumber \\
A(B^{0,+}\to \eta^{\prime} K^{(*)})&=&\frac{\sin\phi}{\sqrt{2}}
A^{0,+}_{q} +\frac{\cos\phi}{\sqrt{2}} A^{0,+}_{s}. \label{eq:amp}
\end{eqnarray}
The decay BRs and CP asymmetries (CPAs) are given by
\begin{eqnarray}
 BR(B^{0,+}\to \eta^{(\prime)} K^{[*]0,+})&=& \frac{G^{2}_{F}|\vec{p}|m^{2}_{B} \tau_{B^{0,+}}}{16\pi
 }\left| A(B^{0,+}\to \eta^{(\prime)}
 K^{[*]0,+})\right|^{2}\,, \\
A_{CP}(B\to \eta^{(\prime)} K^{[*]})&=& { BR(\bar B\to
\eta^{(\prime)} \bar K^{[*]})-BR(B\to \eta^{(\prime)}
 K^{[*]}) \over BR(\bar B\to \eta^{(\prime)} \bar K^{[*]}) + BR(B\to \eta^{(\prime)}
 K^{[*]})}\,,
\end{eqnarray}
which can be evaluated in terms of Eqs.~(\ref{eq:nonlep_n}), (\ref{eq:nonlep_c}) and (\ref{eq:amp}),
where $|\vec{p}|=\sqrt{E^2_{K}-m^2_{K}}$ and
$E_{K}=(m^2_{B}-m^{2}_{\eta^{(\prime)}}+m^2_{K})/2m_B$.

\section{Numerical results and discussions}

\begin{figure}[htbp]
\includegraphics*[width=2.5in]{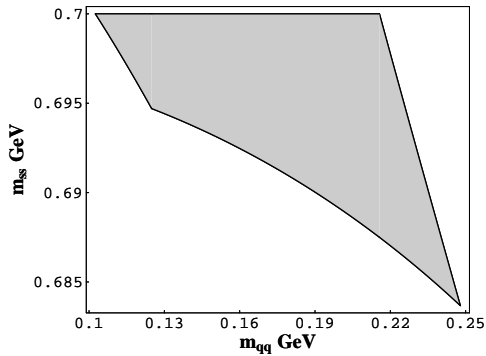}  \caption{The allowed ranges for $m_{qq}$ and $m_{ss}$.}
 \label{fig:bound}
\end{figure}

In the PQCD approach, if we regard the meson wave functions  as
known objects, the remaining unknown theoretical quantities are the
chiral symmetry breaking parameters of states $\eta_{q,s}$ and $K$,
denoted by $m^{0}_{\eta_q,\eta_s,K}$, and the meson decay constants
$f_{B,\eta_{q,s},K}$.
%
 It is known that $f_{K}$ has been
determined quite precisely to be around $0.16$ GeV by experiment,
while the lattice QCD calculations give
$f_{B}=0.216\pm 0.022$ GeV \cite{HPQCD}, which is consistent with
the extracted value from the decay $B^{-}\to \tau \bar\nu_{\tau}$
measured by Belle \cite{Belle}. By low-energy experiments, the decay
constants of $\eta_{q,s}$ are found to be $f_{\eta_q}=(1.07\pm
0.02)f_{\pi}$ and $f_{\eta_s}=(1.34\pm 0.06)f_{\pi}$ \cite{flavor},
respectively. Basically, the undetermined parameters in our
considerations are the parameters $m_{qq}$ and $m_{ss}$. To obtain
the allowed range for $m_{qq,ss}$ in a model-independent way, we
adopt the phenomenological approach. The parameters in Eq.
(\ref{eq:parameters}) are limited to be $\phi=39.3^{\circ}\pm
1.0^{\circ}$, $y=0.81 \pm 0.03$ and $a^2=0.265 \pm 0.010$
\cite{flavor}. With these values, the allowed ranges for $m_{qq}$
and $m_{ss}$ are  presented in Fig.~\ref{fig:bound}.
 From the figure, we find that
$m_{ss}$ has a narrow allowed window around $0.69$ GeV, which can be
understood in terms of the flavor symmetry,  given by
$m_{ss}=\sqrt{2m^{2}_{K}-m^{2}_{\pi}}$ \cite{FK}. However, $m_{qq}$
is relatively broader, given by $0.18\pm 0.08$ GeV.
To do the numerical estimations,
we take
$f_{B}=0.19$ GeV, $f_{\eta_q}=0.14$ GeV and $\phi=39.3^{\circ}$
as the input values.
For the nonperturbative wave functions, we use the results derived by
the LCSR for the light mesons \cite{LCSR}, while for $B$ meson wave function,
we use
 \begin{eqnarray}
   \phi_{B}(x)=N_{B}x^2(1-x)^2\exp\left[-\frac{m^2_{B}x^2}{2\omega_B^2}\right]
   \exp\left[-\frac{\omega_B^2b^2}{2}\right]
 \end{eqnarray}
with $N_{B}=111.2$ and $\omega_B=0.38$ \cite{ChenLi_PRD63}.
 Accordingly, we get the $B\to K$ form factor of $f^{K}_{+}(0)$, defined
in Eq.~(\ref{eq:bpff}),  to be $0.36$.
 From Eq.~(\ref{eq:eta_ff}), we show the form factors
$f^{\eta^{(\prime)}}_{+,T}(0)$
in Table~\ref{tab:value_ff}. From the table, we see clearly that they
will be enhanced with increasing $m_{qq}$.
\begin{table}[hptb]
\caption{ $f^{\eta^{(\prime)}}_{+}(0)$ and
$f^{\eta^{(\prime)}}_{T}(0)$ with three allowed values of
$m_{qq}$.}\label{tab:value_ff}
\begin{ruledtabular}
\begin{tabular}{ccccc}
$m_{qq}$ (GeV) & $f^{\eta}_{+}(0)$ & $f^{\eta}_{T}(0)$ &
$f^{\eta^{\prime}}_{+}(0)$ & $f^{\eta^{\prime}}_{T}(0)$
 \\ \hline
$0.14$ & $0.14$ & $0.14$ & $0.12$ & $0.11$
 \\ \hline
$0.18$ & $0.21$ & $0.20$ & $0.18$ & $0.17$
 \\ \hline
$0.22$ & $0.29$ & $0.29$ & $0.24$ & $0.24$
\end{tabular}
\end{ruledtabular}
\end{table}
In addition, it is easy to understand that the behavior
$f^{\eta}_{+,T}(0)> f^{\eta^{\prime}}_{+,T}(0)$ is always
satisfied as seen from Eq.~(\ref{eq:eta_ff}) 
due to $\cos\phi>\sin\phi$ with $\phi\sim 39.3^{\circ}$. 
This property is different from that in the FSM, given by
\cite{BN_NPB}
\begin{eqnarray}
f^{\eta}_{i}(0)&=&\frac{\cos\phi}{\sqrt{2}} \frac{f_{q}}{f_{\pi}}
f^{\pi}_{i}(0) + \frac{1}{\sqrt{3}} \left( \sqrt{2} \cos\phi\frac{
f_{q}}{f_{\pi}} -  \sin\phi  \frac{f_{s}}{f_{\pi}}\right) f^{\rm
sing}_{i}(0)\, ,\nonumber \\
f^{\eta^{\prime}}_{i}(0)&=&\frac{\sin\phi}{\sqrt{2}}
\frac{f_{q}}{f_{\pi}} f^{\pi}_{i}(0) + \frac{1}{\sqrt{3}} \left(
\sqrt{2} \sin\phi\frac{ f_{q}}{f_{\pi}} +  \cos\phi
\frac{f_{s}}{f_{\pi}}\right) f^{\rm sing}_{i}(0)\,,\label{eq:fs}
\end{eqnarray}
where $f^{\rm sing}_{i}(0)\ (i=+,T)$ correspond to the new form
factors due to the flavor singlet state.
Based on $f^{\pi}_{+}(0)\approx f^{\pi}_{T}(0)\approx 0.26$ calculated by
the LCSRs \cite{LCSR}, we present the numerical results of
Eq.~(\ref{eq:fs}) in Table~\ref{tab:value_fsm}. From the table, we
see  that
$f^{\eta}_{+,T}(0)<f^{\eta^{\prime}}_{+,T}(0)$ in the FSM.
\begin{table}[hptb]
\caption{ $f^{\eta^{(\prime)}}_{+,T}(0)$ with
various values of $f^{\rm
sing}_{i}(0)$ in the FSM.}\label{tab:value_fsm}
\begin{ruledtabular}
\begin{tabular}{ccc}
$f^{\rm sing}_{i}(0)$ & $f^{\eta}_{+,T}(0)$ &
$f^{\eta^{\prime}}_{+,T}(0)$
 \\ \hline
$0.0$ & $0.15$ &  $0.13$
 \\ \hline
$0.1$ & $0.20$ &  $0.25$
 \\ \hline
$0.2$ & $0.25$ &  $0.38$
\end{tabular}
\end{ruledtabular}
\end{table}
Furthermore, by using $|V_{ub}|=3.5\times 10^{-3}$,
$|V_{td}|=8.1\times 10^{-3}$ \cite{PDG06}, Eqs.~(\ref{eq:diffplnu}),
(\ref{eq:difpll}) and (\ref{eq:q2_ff}) and the values in
Tables~\ref{tab:value_ff} and \ref{tab:value_fsm},
we show
the
semileptonic decay BRs
in Table~\ref{table:semirate}. From the table, we find that the results
in both approaches could be consistent with the data of $B^{-}\to
\eta^{(\prime)} \ell \nu_{\ell}$. On the other hand, in our approach, we
always predict $BR(B^{-}\to \eta \ell^{+}
\bar\nu_{\ell})>BR(B^{-}\to \eta^{\prime} \ell^{+} \bar\nu_{\ell})$,
whereas the inequality is reversed in the FSM. Similar
conclusion can be also drawn for the processes of
$B_{d}\to \eta^{(\prime)}\ell^{+}
\ell^{-}$.
We note that the BRs are insensitive to the parametrizations
displayed in Eq.~(\ref{eq:q2_ff}) \cite{CG}.
 \begin{table}[hptb]
\caption{BRs of
 $B^{-}\to \eta^{(\prime)} \ell \bar
\nu_{\ell}$ ( in units of $10^{-4}$) and $\bar B_{d} \to
\eta^{(\prime)} \ell^{+} \ell^{-}$ ( in units of $10^{-7}$) with
$m_{qq}=0.14$, $0.18$ and $0.22$ GeV in our mechanism and
$f^{\rm sing}_{+}(0)=0.0$, $0.1$ and $0.2$ in the FSM.
$\phi=39.3^{\circ}$. }\label{table:semirate}
\begin{ruledtabular}
\begin{tabular}{ccccc}
$m_{qq}$ (GeV) & $B^{-}\to \eta \ell \bar \nu_{\ell}$& $B^{-}\to
\eta^{\prime} \ell \bar \nu_{\ell}$
& $\bar B_{d}\to \eta \ell^{+} \ell^{-}$ & $\bar B_{d}\to \eta^{\prime} \ell^{+} \ell^{-}$ \\ \hline 
$0.14$ & $0.30$ & $0.15 $ & $0.02$ & $0.01$ \\ \hline
$0.18$ & $ 0.67$ & $ 0.35$  & $0.04$ & $0.02$ \\ \hline
$0.22$ & $1.27$ & $0.62$ & $0.07$ &  $0.04$\\\hline\hline $f^{\rm
sing}_{+}(0)$ & $B^{-}\to \eta \ell \bar \nu_{\ell}$& $B^{-}\to
\eta^{\prime} \ell \bar \nu_{\ell}$
& $\bar B_{d}\to \eta \ell^{+} \ell^{-}$ & $\bar B_{d}\to \eta^{\prime} \ell^{+} \ell^{-}$ \\ \hline 
$0.0$ & $0.38$ & $0.18 $ & $0.05$ & $0.03$ \\ \hline
$0.1$ & $ 0.47$ & $ 0.64$  & $0.07$ & $0.10$ \\ \hline
$0.2$ & $0.58$ & $1.39$ & $0.08$ &  $0.21$\\ \hline
Exp & $0.84\pm 0.27 \pm 0.21(<1.4)$ & $0.33\pm 0.60 \pm
0.30(<1.3)$ & $--$ & $--$ \\
 \end{tabular}
\end{ruledtabular}
\end{table}

%
We now give our numerical analysis for the nonleptonic decays $B\to
\eta^{(\prime)} K^{[*]}$. By using the PQCD approach, the values of
factorized and nonfactorized contributions for the  $B$ decays are
shown in Table~\ref{tab:amp_values_n}.
 Based on these values and
 $V_{ts}=-0.041$ and $V_{ub}=4.6\times 10^{-3}e^{-i\phi_3}$
with $\phi_{3}=72^{\circ}$, the predictions for $BR(B\to
\eta^{(\prime)}K^{[*]})$ and $A_{CP}(B\to \eta^{(\prime)}K^{[*]})$
are given in Table~\ref{tab:brk} and
Table~\ref{tab:acpk},
respectively. Our results can be summarized as follows:\\
$\bullet$ From Table~\ref{tab:brk}, we see clearly that with
$m_{qq}=0.22$ GeV, the BRs for $B\to \eta^{(\prime)} K^{[*]}$
are consistent with the WA data.
It is interesting to note that
by increasing $m_{qq}$, $BR(B\to \eta^{(')} K)$ tend to be small (large),
while $BR(B\to \eta^{(')} K^{*})$ to be large (small),
favored by the experiments.
\\
$\bullet$
 As seen from Table~\ref{tab:brk}, with the same value of $m_{qq}$,
 $BR(B\to \eta K)<O(10^{-1})
 BR(B\to \eta^{\prime} K)$, while $BR(B\to \eta K^{*})>
 BR(B\to \eta^{\prime} K^{*})$.
 The phenomena could be ascribed to the signs in the amplitudes of $B\to (\eta_q,\, \eta_s)K^{(*)}$ by comparing Eqs. (\ref{eq:nonlep_n}),
 (\ref{eq:nonlep_c}) and (\ref{eq:amp}) with 
 the specific values of
    $F^{0,+}_{Pa}$ and $F^{0,+}_{P(b+c)}$ in
   Table~\ref{tab:amp_values_n}. 
\\
$\bullet$
   From Table~\ref{tab:acpk}, we find that for
$m_{qq}=0.22$ GeV  $A_{CP}(B_u\to \eta K^+)$  is as large as
$-30\%$, which agrees well with the data, whereas the other two sets
of $m_{qq}$ lead to positive and small asymmetries.
In addition, our prediction for $A_{CP}(B_{d}\to \eta K^{*0})$ is
too small, while that of $A_{CP}(B_{u}\to \eta K^{*+})$ is too
large, in comparison with the data. If future experiments display
the current tendencies for these CPAs, such phenomena will become
new puzzles. 

%
\begin{table}[hptb]
\caption{ Factorizable and nonfactorizable parts for the decays
$B\to \eta_{q,s}K^{[*]}$ with $m_{qq}=0.22$ GeV, where the values in
the square brackets are for $B\to \eta_{q,s}K^{*}$.}
\label{tab:amp_values_n}
\begin{ruledtabular}
\begin{tabular}{cccccc}
$F^{0}_{Pa}10^{2}$ & $N^{0}_{Pa}10^{5}$ & $F^{0}_{P(b+c)}10^{2}$ &
$N^{0}_{P(b+c)}10^{4}$ & $F^{0}_{Pc}10^{2}$ & $N^{0}_{Pc}10^{4}$
 \\ \hline
$-1.10$ & $6.36-i2.06$ & $-0.55$ & $-0.33+i 1.22$ & $0.26$ &
$-7.17+i 3.39$
\\
$[-0.42]$ & $[3.89-i2.37]$ & $[0.45]$ & $[-0.89+i1.74]$ & $[0.19]$ &
$[-7.88+i 2.56]$
\\\hline\hline
$F^{0}_{Pd}10^{3}$ & $N^{0}_{Pd}10^{5}$ & $F^{0}_{Pe}10^{3}$ &
$N^{0}_{Pe}10^{5}$ & $F^{0}_{Tf}10^{2}$ & $N^{0}_{Tf}10^{3}$
\\\hline
$-0.61+i 2.43 $ & $-5.77-i 9.62$ & $-0.44 + i 1.25$ &
$-0.51-i4.62$ & $-0.61$ & $3.61 -i 1.57$
 \\
$[-0.19+i2.37]$ & $[-5.05-i3.36]$ & $[0.30-i1.86]$ & $[-5.02-i9.06]$
& $[-0.41]$ & $[4.00-i1.29]$
\\\hline\hline
$F^{+}_{Pa}10^{2}$ & $N^{+}_{Pa}10^{5}$ & $F^{+}_{P(b+c)}10^{2}$ &
$N^{+}_{P(b+c)}10^{4}$ & $F^{+}_{Pc}10^{2}$ & $N^{+}_{Pc}10^{4}$
 \\ \hline
$-1.05$ & $3.55-i0.27$ & $-0.54$ & $-3.56+i 1.71$ & $0.21$ &
$-6.24+i 1.70$
\\
$[-0.43]$ & $[-1.86-i2.61]$ & $[0.45]$ & $[-0.89+i1.74]$ & $[0.188]$ & $[-7.64+i3.53]$\\
\hline\hline $F^{+}_{Pd}10^{3}$ & $N^{+}_{Pd}10^{5}$ & $F^{+}_{Pe}10^{3}$
& $N^{+}_{Pe}10^{5}$ & $F^{+}_{Tf}10^{2}$ & $N^{+}_{Tf}10^{3}$
 \\ \hline
$-0.63+i 2.20 $ & $-2.86-i 4.46$ & $-0.50 + i 1.60$ &
$-1.59-i2.98$ & $-0.45$ & $3.27 -i 0.90$
 \\
$[-0.09+i2.37]$ & $[-2.21-i0.72]$ & $[0.29-i1.83]$ & $[-2.83-4.07]$
& $[-0.41]$ & $[3.85-i 1.74]$
 \\\hline\hline
$F^{+}_{Tg}10^{2}$ & $N^{+}_{Tg}10^{3}$ & $F^{+}_{Th}10^{3}$ &
$N^{+}_{Th}10^{3}$ & $F^{+}_{Ti}10^{3}$ & $N^{+}_{Ti}10^{3}$
 \\\hline
 $10.03$ & $-1.16 + i 0.18$ & $2.38+i0.02$ &
$0.99 +i 1.38$ & $-1.02-i0.02$ & $0.27 + i 1.13$
 \\
$[11.60]$ & $[-1.55+i0.06]$ & $[-2.19-i1.14]$ & $[1.09+i0.70]$ &
$[1.61+i 0.95]$ & $[0.89+i1.55]$
 \\
\end{tabular}
\end{ruledtabular}
\end{table}
\begin{table}[hptb]
\caption{$BR(B\to \eta^{(\prime)}K^{[*]})$ (in units of $10^{-6}$)
 with $m_{qq}=0.14$, $0.18$ and $0.22$ GeV
 as well as the
  world average (WA) values \cite{HFAG}.}\label{tab:brk}
\begin{ruledtabular}
\begin{tabular}{ccccc}
$m_{qq}$ & $B_{d}\to \eta K^0$ & $B_{d}\to \eta^{\prime} K^0$ &
$B_{u}\to \eta K^{+}$ & $B_{u}\to \eta^{\prime} K^{+}$
 \\ \hline
$0.14$ & $3.01$ & $31.44$ & $5.66$ & $34.60$
\\ \hline
$0.18$ & $0.28$ & $44.01$ & $1.26$ & $47.36$
 \\ \hline
$0.22$ & $1.43$ & $62.69$ & $1.52$ & $65.04$
 \\ \hline
 WA & $<1.9$ & $64.9 \pm 3.5$ & $2.2\pm 0.3 $ &
$69.7^{+2.8}_{-2.7}$
\\\hline\hline
$m_{qq}$ & $B_{d}\to \eta K^{*0}$ & $B_{d}\to \eta^{\prime} K^{*0}$
& $B_{u}\to \eta K^{*+}$ & $B_{u}\to \eta^{\prime} K^{*+}$
 \\ \hline
$0.14$ & $11.54$ & $8.21$ & $11.74$ & $10.06$
\\ \hline
$0.18$ & $15.91$ & $5.76$ & $15.94$ & $8.12$
 \\ \hline
$0.22$ & $22.31$ & $3.35$ & $22.13$ & $6.38$
 \\ \hline
 WA & $16.1\pm 1.0$ & $3.8\pm 1.2$ & $19.5^{+1.6}_{-1.5}$ & $4.9^{+2.1}_{-1.9}$
\end{tabular}
\end{ruledtabular}
\end{table}
\begin{table}[hptb]
\caption{ $A_{CP}(B\to
\eta^{(\prime)}K^{[*]})$ (in unit of $10^{-2}$)
  with $m_{qq}=0.14$, $0.18$ and $0.22$ GeV as well as the
  world average (WA) values \cite{HFAG}.}\label{tab:acpk}
\begin{ruledtabular}
\begin{tabular}{ccccc}
$m_{qq}$ & $B_{d}\to \eta K^0$ & $B_{d}\to \eta^{\prime} K^0$ &
$B_{u}\to \eta K^{+}$ & $B_{u}\to \eta^{\prime} K^{+}$
 \\ \hline
$0.14$ & $-2.10$ & $0.69$ & $5.62$ & $-5.28$
\\ \hline
$0.18$ & $-2.47$ & $0.57$ & $5.88$ & $-6.19$
 \\ \hline
$0.22$ & $4.41$ & $0.48$ & $-30.64$ & $-6.88$
 \\ \hline
WA & $--$ & $--$ & $-29\pm 11$ & $3.1 \pm 2.1$
\\ \hline\hline
$m_{qq}$ & $B_{d}\to \eta K^{*0}$ & $B_{d}\to \eta^{\prime} K^{*0}$
& $B_{u}\to \eta K^{*+}$ & $B_{u}\to \eta^{\prime} K^{*+}$
 \\ \hline
$0.14$ & $0.79$ & $-0.82$ & $-15.79$ & $8.39$
\\ \hline
$0.18$ & $0.67$ & $-0.98$ & $-20.51$ & $8.83$
 \\ \hline
$0.22$ & $0.57$ & $-1.30$ & $-24.57$ & $4.60$
 \\ \hline
WA & $19\pm 5$ & $-8\pm 25$ & $2\pm 6$ & $30^{+33}_{-37}$
\end{tabular}
\end{ruledtabular}
\end{table}

Finally, we remark that in
the quark-flavor scheme,
 as the errors in the decay constants of $f_{q}$ and $f_{s}$
are only $2\%$ and $4\%$, respectively, their effects on 
BRs and CPAs are mild. However, the influence from the mixing angle 
$\phi$ could be larger.
We present the results with the error of $\phi$
in Table~\ref{tab:br-acp}.
\begin{table}[hptb]
\caption{BRs (in units of $10^{-6}$)
 and CPAs (in units of $10^{-2}$) for $B\to \eta^{(\prime)}K^{[*]}$ decays with
 $m_{qq}=0.22$ GeV and $\phi=39.3^{\circ} \pm 1.0^{\circ}$.}\label{tab:br-acp}
\begin{ruledtabular}
\begin{tabular}{ccccc}
Obs. & $B_{d}\to \eta K^0$ & $B_{d}\to \eta^{\prime} K^0$ &
$B_{u}\to \eta K^{+}$ & $B_{u}\to \eta^{\prime} K^{+}$
 \\ \hline
BR & $1.43^{+0.34}_{-0.31}$ & $62.69^{+0.30}_{-0.34}$ & $1.52\pm^{+0.16}_{-0.13} $ & $65.04^{+ 0.12}_{-0.15}$ \\
$A_{CP}$ & $4.41^{+0.57}_{-0.44}$ & $0.48\pm 0.009 $ &
$-30.64^{+4.12}_{-2.87}$ & $-6.88^{+0.13}_{-0.12}$
\\\hline\hline
Obs. & $B_{d}\to \eta K^{*0}$ & $B_{d}\to \eta^{\prime} K^{*0}$ &
$B_{u}\to \eta K^{*+}$ & $B_{u}\to \eta^{\prime} K^{*+}$
 \\ \hline
BR & $22.31^{+0.28}_{-0.29}$ & $3.35^{+0.29}_{-0.27}$ & $22.13^{+0.26}_{-0.27}$ & $6.38\pm 0.26$\\
$A_{CP}$ & $0.57\pm 0.011$ & $-1.30\pm 0.08$ &
$-24.57^{+0.72}_{-0.27}$ & $4.60^{+1.16}_{-1.32}$

\end{tabular}
\end{ruledtabular}
\end{table}


\section{Conclusions}

Due to the current experimental limits on the mixing parameters of
the $\eta$ and $\eta^{\prime}$ mesons, we have studied the
phenomenologically allowed ranges for $m_{ss}$ and $m_{qq}$.
Explicitly, we have
 found that
$m_{ss}$ is around 0.69 GeV and $m_{qq}=0.18\pm 0.08$ GeV. We have
shown that the semileptonic decays of $B^{-}\to \eta^{(\prime)} \ell
\bar\nu_{\ell}$ are sensitive to $m_{qq}$
 and thus they can provide
strong constraints on its value. In addition, our mechanism based on
the quark-flavor mixing scheme naturally leads to $f^{ \eta^{\prime}}_{+}(0) <
f^{ \eta}_{+}(0)$
as well as
$BR(B^-\to \eta \ell^{-}
\bar\nu_{\ell})>BR(B^{-}\to \eta^{\prime} \ell^{-} \bar\nu_{\ell})$,
in contrast with the reversed inequalities
in the FSM due to the flavor-singlet contribution
\cite{CG,KOY}.
Similar conclusions can also be drawn for
 the decays
$B_{d}\to \eta^{(\prime)} \ell^{+} \ell^{-}$.
It is interesting to note that the future measurements on $BR(B^{-}\to
\eta^{(\prime)}\ell \bar\nu_{\ell})$ and $BR(B_{d}\to \eta^{(\prime)}
\ell^{+} \ell^{-})$
can be used to distinguish the two flavor mechanisms. Moreover, we
have shown that  $BR(B\to \eta^{(\prime)} X)$ with $X=(\ell^{-}
\bar\nu_{\ell},\; \ell^{+} \ell^{-})$ are enhanced and in
particular, the puzzle of the large $BR(B\to \eta^{\prime} K)$ can
be solved with a reasonable large value of $m_{qq}$. We have also
demonstrated that $A_{CP(}B^{\pm}\to \eta K^{\pm})$ can be as large
as $-30\%$ and $BR(B\to \eta^{(\prime)}K^*)$ are consistent with the
current data. Finally, we remark that our results for $A_{CP}(B\to
\eta K^*)$ do not agree with the experimental values. According to
our analysis, currently, they are the most incomprehensible
phenomena. Other mechanisms as well as more precise measurements are
needed for a complete description of all the above decays.
\\

{\bf Acknowledgments}\\

The authors would like to thank  Prof. Hai-Yang Cheng and Prof.
Hsiang-Nan Li for useful discussions. This work is supported in part
by the National Science Council of R.O.C. under Grant
 \#s:NSC-95-2112-M-006-013-MY2 and NSC-95-2112-M-007-059-MY3.

\end{document}